%% file: miglio_jenam05CoAst.tex
\documentclass[mypaper,8pt,twoside]{CoAst}
\usepackage{epsf,graphicx,fancyhdr}
\input{CoAst_layo}

\begin{document}
\sf

\chapterDSSN{Effects of ``extra-mixing'' processes on the periods of high-order gravity modes in main-sequence stars}{A. Miglio, J. Montalb\'an and A. Noels}

\Authors{A. Miglio, J. Montalb\'an, A. Noels} 
\Address{Institut d'Astrophysique et de G\'eophysique de l'Universit\'e de Li\`ege,
All\'ee du 6 Ao\^ut, 17 B-4000 Li\`ege, Belgium}

\noindent
\begin{abstract}
In main-sequence stars, the chemical composition gradient that develops at the edge of the convective core is responsible for a non-uniform period spacing of high-order gravity modes.
In this work we investigate, in the case of a 1.6 $\rm M_\odot$ star, the effects on the period-spacing of extra mixing processes in the core (such as diffusion and overshooting).
\end{abstract}

\section{Effects of overshooting and diffusion on the central $\mu$-profile}\label{miglio-sec1}
%figura
\figureDSSN{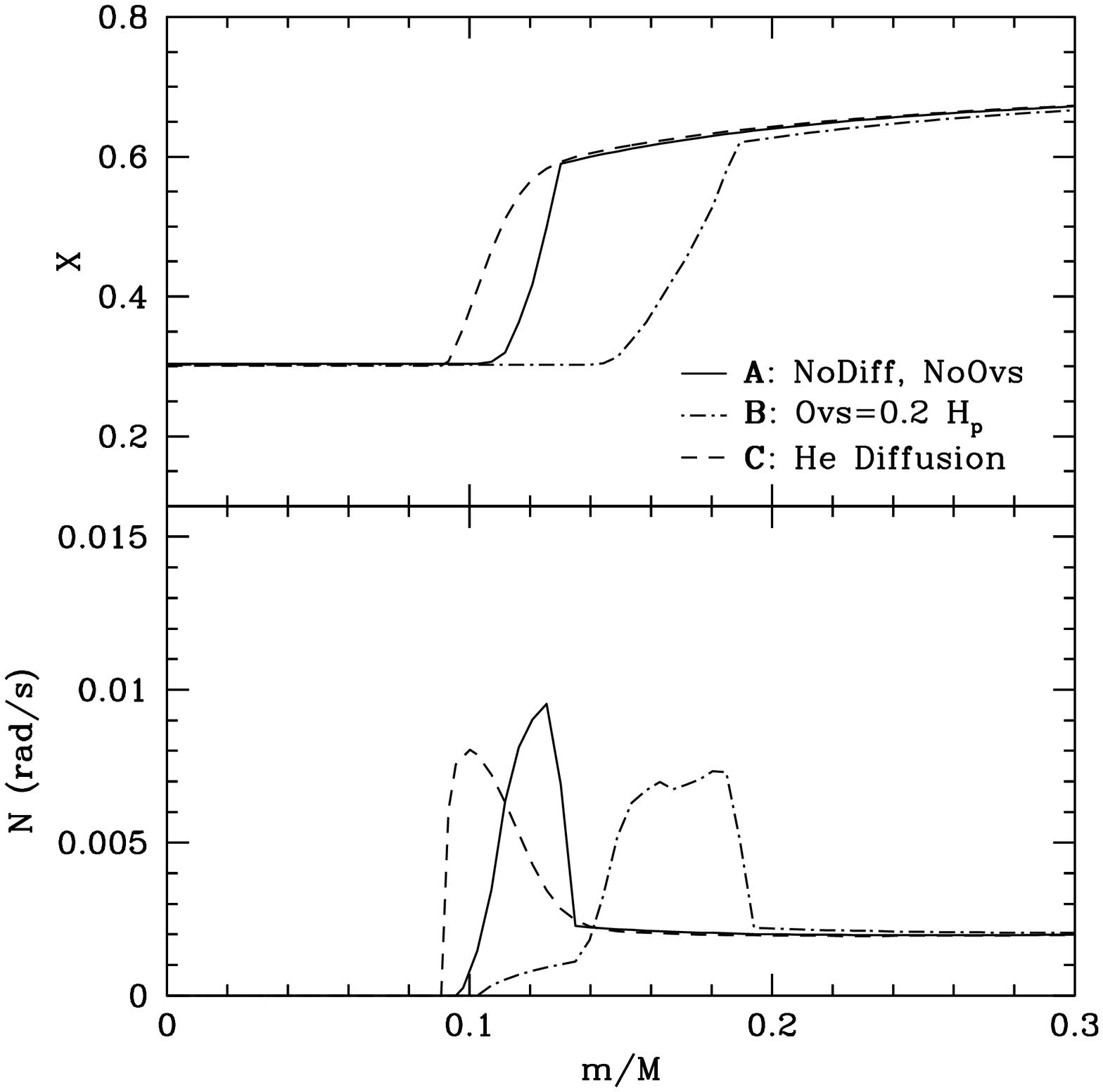}
{Behavior of the the hydrogen abundance profile (upper panel) and of the  Brunt-V\"ais\"al\"a frequency (lower panel) in models of 1.6 $\rm M_\odot$ with $X_c\simeq 0.3$. The different lines correspond to models calculated with no extra-mixing (A, continuous lines), overshooting (B, dashed-dotted) and helium diffusion (C, dashed). The different location and sharpness of the chemical composition gradient determines the behaviour of $N$ (lower panel).}{fig:profiles}{ht!}{clip,angle=0,width=0.7\textwidth }  
We consider three models of a 1.6 $\rm M_\odot$ star on the main sequence: a model without any ``extra-mixing'' process (model A),  considering overshooting from the convective core (B) and including helium diffusion (C).
Though the  central hydrogen abundance is the same  ($X_c=0.3$), the models have a different chemical composition profile near the outer edge of the convective core (see Fig. \ref{fig:profiles}). Taking A as the reference model, we see that overshooting (B) displaces the location of the $\mu$-gradient, whereas diffusion (C) leads to a smoother chemical composition profile. As shown in the lower panel of Fig. \ref{fig:profiles}, such differences are also reflected in the behaviour of the Brunt-V\"ais\"al\"a frequency $N$ that, in its turn, determines the properties of gravity modes.

\section{Effects on the period spacing}
In white dwarfs it has been theoretically predicted and then observed (see for instance Metcalfe et al. (2003) and references therein) that the period spacing (hereafter $\Delta P$) of g-modes is not constant, contrary to what is predicted by the first order asymptotic approximation of gravity modes (Tassoul 1980). This has been interpreted as the signature of sharp variations in $N$ due to chemical composition gradients in the envelope and core of the star.

In analogy with the case of white dwarfs, in main-sequence models with a convective core we expect the formation of a {\it nonuniform period distribution}; this is in fact the case as presented in Fig. \ref{fig:gmodes}. The period spacing shows clear periodic components superposed to a constant $\Delta P$ expected for a model without sharp variations in $N$.
The periodicity and amplitude of these components can be related to the location and sharpness of the $\mu$-gradient region by means of analytical expressions.

\subsection{Analytical approximations}
As described e.g. in Montgomery et al. (2003) the effect of a sharp feature in the model (a chemical composition gradient, for instance) can be estimated as the periodic component of the difference $\delta P$ between the periods of the star showing such a sharp variation and the periods of an otherwise fictitious smooth model.

As a first example we model as a {\it step function} the sharp feature $\delta N$ in the Brunt-V\"ais\"al\"a frequency located at a normalized radius $x=x_\mu$.

We then define
\begin{equation}
\Pi^{-1}(x)=\int_{x_0}^x{\frac{|N|}{x'}dx'}\;,  \;\Pi_0^{-1}=\int_{x_0}^1{\frac{|N|}{x'}dx'}\; \; {\rm and}\; \;
\Pi_\mu^{-1}=\int_{x_0}^{x_\mu}{\frac{|N|}{x'}dx'}
\end{equation}
where $x_0$ is the boundary of the convective core and we consider a model with a radiative envelope.

\figureDSSN{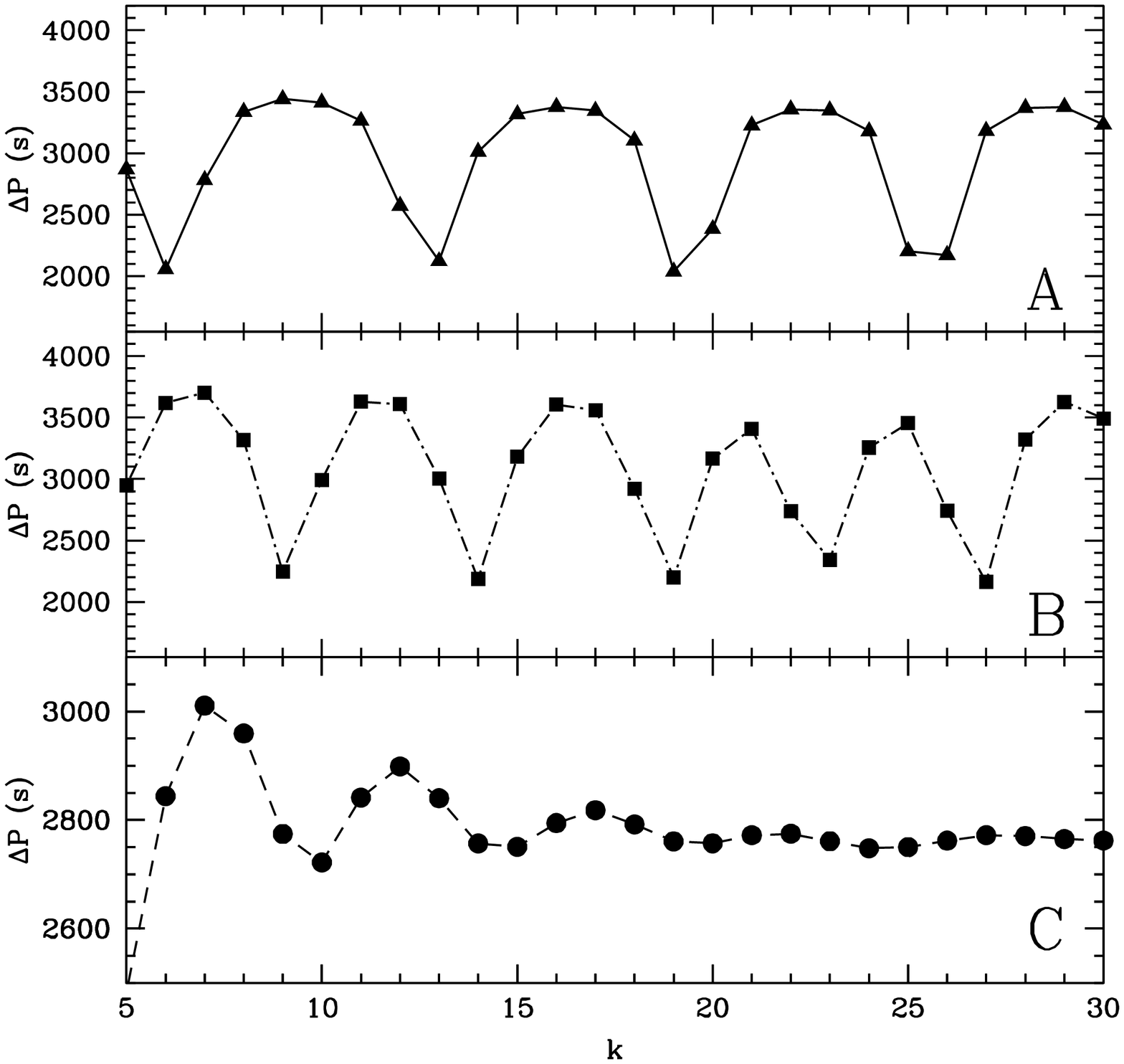}{Period spacing $\Delta P=P_{k+1}-P_k$ as a function of the radial order $k$ in $\ell=1$ g-modes for the models presented in Figures \ref{fig:profiles} and \ref{fig:bv}. The periods of g modes were computed using the adiabatic stellar oscillation code OSC.}{fig:gmodes}{!ht}{clip,angle=0,width=0.7\textwidth}

Following the approach of  Montgomery et al. (2003), and using the asymptotic expression for g-modes periods $P_k$ derived by Tassoul (1980):
\begin{equation}P_k=\pi^2\frac{\Pi_0}{L} \left(2k+n_e\right)
\label{eq:tassoul}
\end{equation}
where $n_e$ is the effective polytropic index of the superficial layers, $k$ is the radial order of the mode and $L=[\ell(\ell+1)]^{1/2}$, we find
\begin{equation}
\delta P_k\propto\frac{\Pi_0}{L} A \cos{\left(2\pi\,\frac{\Pi_0}{\Pi_\mu} k+\phi \right)}{\rm,}
\label{eq:varia}
\end{equation}
where $A$ is related to the sharpness of the variation in $N$.

From this simple approach we derive that the signature of a sharp feature in the  Brunt-V\"ais\"al\"a frequency is a {\it periodic component in the periods of oscillations}, and therefore  in the period spacing $\Delta P$, whose periodicity in terms of the radial order $k$ is given by
\begin{equation}
\Delta k\simeq\frac{\Pi_\mu}{\Pi_0}
\label{eq:variak}
\end{equation}
and whose amplitude does not depend on the order $k$.

We now compare this approximation to the numerically computed period spacing. In Fig.~\ref{fig:gmodes} the periods (in terms of $k$) of the components are approximately 7 for model A and 5 for model B and C. Following Eq. (\ref{eq:variak}) these periods should correspond to a location of the discontinuity (expressed as $\Pi_0/\Pi_\mu \simeq k^{-1}$) of ~0.14 and ~0.2: as shown in Fig. \ref{fig:bv}, these estimates  accurately describe the locations of the sharp variation of $N$ in the models.

The period spacing of the model computed with diffusion deserves, however, further inspection. As shown in the lower panel of Fig. \ref{fig:gmodes}, the amplitude of the components in the period spacing of model C, compared to model A and B, is considerably reduced. Moreover, the amplitude also becomes a decreasing function of the order $k$: this behaviour can be directly related to a smoother chemical composition profile.

In fact, the simple approach followed so far allows us to easily evaluate the effect of having a smoother variation in the Brunt-V\"ais\"al\"a frequency. Instead of modelling $\delta N$ as a step function, we use a {\it ramp function} that, as shown in Fig. \ref{fig:bv}, better represents the variation of $\delta N$ in model C.

In this case integration by parts leads to a sinusoidal component in $\delta P_k$ whose {\it amplitude is modulated by a factor $1/P_k$} and therefore decreases with increasing $k$, i.e.
\begin{equation}
\delta P_k\propto\frac{\Pi_0}{L}\frac{A'}{P_k}\cos{\left(2\pi\,\frac{\Pi_0}{\Pi_\mu} k+\phi' \right)}\;.
\end{equation}

\figureDSSN{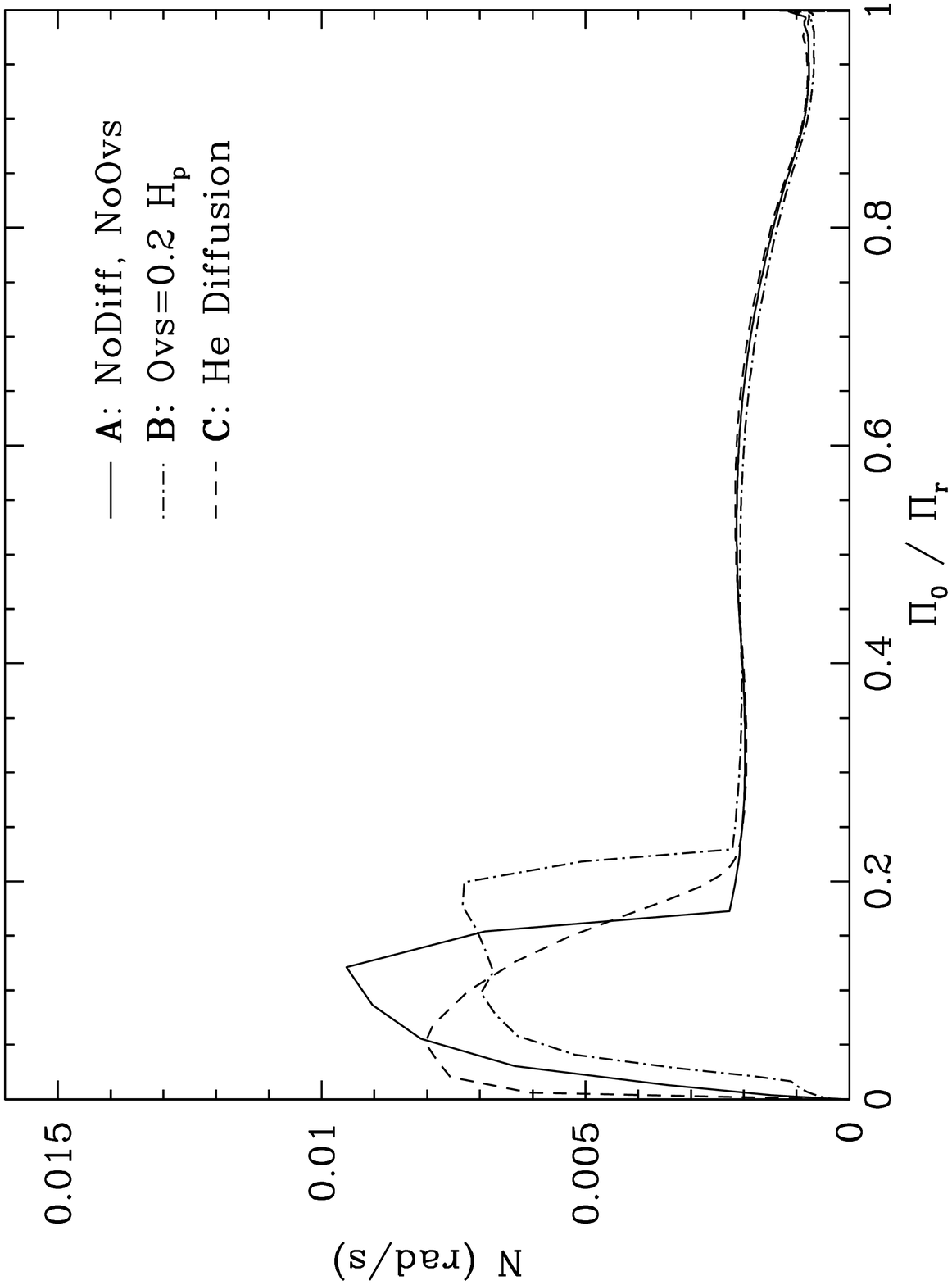}{ The Brunt-V\"ais\"al\"a frequency versus $\Pi_\mu/\Pi_0$ for the models. Whereas the sharp variation in $N$ in model A and B can be represented by a step function, in the case of model C (calculated with diffusion) it is better modelled  by a ramp function.}{fig:bv}{!ht}{clip,angle=-90,width=0.6\textwidth}

This simplified approach is therefore sufficient to account for the behaviour  of $\Delta P$ (Fig. \ref{fig:gmodes}) in the model computed with diffusion, where the sharp feature in $N$ described by a discontinuity not in $N$ itself, but in its first derivative, generates a periodic component whose amplitude decreases with $k$.
\section{Conclusions and prospects}
In main-sequence stars, similarly to the case of white dwarfs, the deviations from a constant g-mode period spacing are sensitive probes of $\mu$-gradients that develops at the outer edge of a convective core.
These deviations can be interpreted by means of simple analytical expressions that could represent a possible seismic tool to study the detailed properties of chemical mixing in $\gamma$ Doradus and SPB stars, where high-order gravity modes are observed.

The question whether such signatures could be detected given realistic observational errors and other uncertainties (e.g. effects of rotation on g-modes periods)  needs however further investigation. A more detailed study will be presented in a forthcoming paper.

\acknowledgments{A.M. and J.M. acknowledge financial support from the Prodex-ESA Contract 15448/01/NL/Sfe(IC).
}

\References{
Metcalfe T. S., Montgomery, M. H., \& Kawaler, S. D. 2003, MNRAS 344, L88\\
Montgomery, M. H., Metcalfe T. S., \& Winget, D. E. 2003, MNRAS 344, 657\\
Tassoul, M. 1980 ApJS 43, 469
}

\end{document}

%% file: CoAst_layo.tex
\renewcommand{\chaptermark}[1]{\markboth{#1}{}}

\renewcommand{\headrulewidth}{0pt}
\newcommand{\kopf}{\small\itshape Comm. in Asteroseismology\\ Vol. 144, 2003}
\newcommand{\Authors}[1]{\begin{center}\normalsize\bf\sf #1 \end{center}}

\renewcommand{\author}[1]{\begin{center}\normalsize\bf\sf #1 \end{center}}
\newcommand{\Address}[1]{\begin{center}\small\sf #1 \end{center}}

\renewenvironment{abstract}{\section*{Abstract}\normalsize\sf}{}
\newcommand{\References}[1]{\begin{flushleft}{\large References\\}\vspace*{2mm}\small #1 \end{flushleft}}

\newcommand{\chapterDSSN}[2]{\chapter[\sf\normalsize #1\\ \footnotesize \hspace*{5mm}by #2 \sf\normalsize][]{#1\\}\rhead[\fancyplain{}{\sf\footnotesize \center{#1}}]{\fancyplain{}{\sffamily\thepage}}\lhead[\fancyplain{\kopf}{\sffamily\thepage}]{\fancyplain{\kopf}{\sf\footnotesize \center{#2}}}}

\newcommand{\figureDSSN}[5]{\begin{figure}[#4]
\centering
\includegraphics*[#5]{#1}
\caption{#2}
\label{#3}
\end{figure}}

\renewcommand{\chaptername}{}
\newcommand{\acknowledgments}[1]{\vspace*{5mm}\noindent\begin{bf}Acknowledgments. \end{bf} #1}